\documentstyle[preprint,floats,pra,eqsecnum,aps,psfig]{revtex}
\tightenlines
\begin{document} \draft

\title{\large \bf Lorentz Covariance and Internal Space-time Symmetry of
Relativistic Extended Particles}

\author{Y. S. Kim}
\address{Department of Physics, University of Maryland, College Park,
Maryland 20742}

\maketitle

\begin{abstract}
The difference between Lorentz invariance and Lorentz covariance is
discussed in detail.  A covariant formalism is developed for
the internal space-time symmetry of extended particles, especially
in connection with the insightful observations Feynman made during the
period 1969-72.  A Lorentz-group formalism is presented for the harmonic
oscillator model of Feynman, Kislinger and Ravndal, which was originally
based on hadronic mass spectra.  This covariant version allows us to
construct a parton distribution function by Lorentz-boosting the
oscillator wave function of a hadron at rest.  The role of the
time-separation variable is discussed in detail.  It is shown that,
due to our inability to make measurements on this variable, it belongs
to Feynman's rest of the universe.  Our failure to observe the rest of
the universe leads to an increase in entropy.
\end{abstract}


\section{Introduction}\label{intro}
Ever since the present form of quantum mechanics was formulated in 1927,
the most pressing problem has been and still is how to combine quantum
mechanics with special relativity.  During this process, the success of
quantum electrodynamics led many physicists to believe that quantum field
theory is the ultimate answer to this question.  On the other hand, in
field theory, we produce measurable numbers by calculating scattering
matrix elements using Feynman diagrams, where initial and final particles
are all free particles.  Indeed, the present form of quantum field
theory, although covariant, can accommodate only scattering problems.

How about bound states?  The calculation of the Lamb shift is one of the
triumphs of quantum field theory.  In order to calculate this shift, we
need wave functions of the hydrogen atom whose behavior at the origin
depends on their angular momentum states.  Can we then calculate the
hydrogen wave function and the Rydberg formula within the framework of
quantum field theory?  The answer to this question is clearly {\it No}.
The Rydberg levels come from the localization condition for probability
distribution.  Does field theory say anything about localized
probability distributions?  The answer to this question is again
{\it No}.

The hydrogen atom appears as a localized probability entity when it is at
rest.  Its angular momentum state determines how the distribution will
appear to the observer in a rotated coordinate system.  This is of course
one of the fundamental problems in atomic spectra~\cite{wig59}.  We can
now consider how the wave function would look to observes in different
Lorentz frames.  This had remained only as an academic question until
Hofstadter~\cite{hofsta55} observed in 1955 that the proton is not a
point particle, but its charge has a distribution.  Hofstadter detected
this from the relativistic recoil effects in the scattering of electrons
by protons.

In spite of many laudable efforts to explain this form factor within
the framework of quantum field theory, the workable model for the form
factors did not emerge until after Gell-Mann's formulation of the
quark model, in which all hadrons are bound states of quarks and/or
anti-quarks~\cite{gell64}.  The question then is whether we can use
the existing models of quantum mechanics, such as the nuclear shell
model, to explain hadronic mass spectra~\cite{owg64}.  For the mass
spectra, one of the most effective models has been and still is the
model based on harmonic oscillator wave functions~\cite{owg64,fkr71}.
The basic advantage of the oscillator model is that its mathematics is
simple and transparent, and it does not bury physics in mathematics
even though it does not always produce the most accurate numerical
results.  The oscillator model will prevail if there are no other
models capable of producing better numerical results.  This appears
to be the present status of the quark model.

In the quark model, the charge distribution within the proton comes
from the distribution of the charged particles inside the hadron.
The success of the oscillator model for static or slow-moving
hadrons does not necessarily mean that the model can be extended to
the relativistic regime.  The calculation of the form factor
with Gaussian wave functions results in an exponential decrease for
large momentum-transfer variables.  However, this wrong behavior
comes from the use of non-relativistic wave functions for relativistic
problems.  Feynman {\it et al.} made an attempt to construct a
covariant oscillator model~\cite{fkr71}.  Even though they did not
achieve this goal in their paper, Feynman {\it et al.} quote the work
of Fujimura {\it et al.}~\cite{fuji70} who calculated the nucleon form
factor by taking into account the effect of the Lorentz-squeeze on the
oscillator wave functions.

After studying these original papers, we can raise our level of
abstraction.  We observe first that the spherical harmonics can
represent the three-dimensional rotation group, while serving as wave
functions for the angular variables.  Then, we can ask whether there
are wave functions which can represent the Poincar\'e group.  We can
specifically ask whether it is possible to construct a set of
normalizable harmonic oscillator wave functions to represent the
Poincar\'e group.  If {\it Yes}, the wave functions can be
Lorentz-boosted.  These wave functions then have to go through another
set of tests.  Are they consistent with the existing laws of quantum
mechanics.  If {\it Yes}, they then have to be exposed to the most
cruel test in physics.  Do they explain what we observe in high-energy
laboratories?

The purpose of this paper is to show that we can use the oscillator
wave functions to answer the question of whether quarks are partons.
While the quark model is valid for static hadrons, Feynman's parton
picture works only in the Lorentz frame where the hadronic speed is
close to that of light~\cite{fey69}.  The quark model appears to be
quite different from the parton model.  On the other hand, they are
valid in two different Lorentz frames.  The basic question is whether
the quark picture and the parton picture are two different
manifestations of the same covariant entity.  We shall discuss first
the internal space-time symmetries of relativistic particles in terms
of appropriate representations of the Poincar\'e group~\cite{wig39}.
We then construct the oscillator wave functions satisfying the
above-mentioned theoretical
criterions~\cite{dir45,yuka53,markov56,kno79jm,knp86}.
This oscillator formalism will explains both the quark and the parton
pictures in two separate Lorentz frames.  This formalism produces all
the peculiarities of Feynman's original form of the parton picture
including the incoherence of parton cross sections.

In addition, Feynman's parton picture raises an important question
in measurement theory.  Let us consider again a hadron consisting of
two quarks.  Then there is a Bohr-like radius measuring the space-like
separation between the quarks.  There is also a time-like separation
between the quarks, and this variable becomes mixed with the
longitudinal spatial separation as the hadron moves with a relativistic
speed~\cite{dir27}.  There are at present no quantum measurement
theories to deal with the above-mentioned time-like separation.
Because we do not know, we do not mention it.

We can afford to do this in nonrelativistic quantum mechanics.  However,
in the relativistic regime, the time-separation variable becomes as
important as the longitudinal separation because they become mixed when
the system is Lorentz-boosted.  Can we still pretend not to know
anything about this variable?  The best solution to this problem is
to put this variable into Feynman's rest of the
universe~\cite{fey72,hkn98}.  The net result is then that this
unobservable variable leads to an increase in entropy.

For many years, the author has been trying to discuss these problems
with his colleagues, but he has to confess that there has been a
difficulty in communication.  The difficulty seems to come from the
time-separation variable which does not exist in the present form of
non-relativistic quantum mechanics, but which plays the pivotal role
in the relativistic world.  As for the communication problem, the
fundamental issue appears to be the difference between Lorentz
invariance and Lorentz covariance.   This difference is not clearly
understood by most physicists these days.  For this reason, we start
this paper with a discussion of this problem.

In Sec.~\ref{kant}, we discuss the historical origin of the concept of
covariance.  It is pointed out that, in formulating the theory of
relativity, Immanuel Kant played the pivotal role before Einstein.
Kant's cultural background is also discussed in detail.  We explain
why the author is in a better position to appreciate the Kantian view
of the world.
In Sec.~\ref{littleg}, it is noted that relativistic particles have
their internal space-time structures.  It was Eugene Wigner who
formulated this problem in terms of the subgroups of the Poincar\'e
group known as the little groups~\cite{wig39}.  In this section,
we present a brief history of applications of the little groups to
internal space-time symmetries of relativistic particles.
In Sec.~\ref{covham}, we construct representations of the little
group for massive particles using harmonic-oscillator wave functions.
In Sec.~\ref{parton}, it is shown that the Lorentz-boosted oscillator
wave functions exhibit the peculiarities of Feynman's parton model in
the infinite-momentum limit.  In Sec. \ref{rest}, we point out first
that Feynman's parton picture contains Feynman's rest of the universe.
It is then shown that our failure in measuring the time separation
variable leads to an increase in entropy.

Much of the concept of Lorentz-squeezed wave function is derived from
elliptic deformation of a sphere resulting in a mathematical technique
group called contractions~\cite{inonu53}.  In Appendix~\ref{o3e2}, we
discuss the contraction of the three-dimensional rotation group to the
two-dimensional Euclidean group.  In Appendix~\ref{contrac}, we
discuss the little group for a massless particle as the
infinite-momentum/zero-mass limit of the little group for a massive
particle.

\section{Covariance and Its Historical Background}\label{kant}
The words ``Lorentz invariance'' and ``Lorentz covariance'' are
frequently used in physics.  Are they the same word or are they
different?  Why is this question so serious?  Unlike classical physics,
modern physics depends heavily on observer's state of mind or
environment.  In special relativity, observers in different Lorentz
frames see the same physical system differently.  The importance of the
observer's subjective viewpoint was emphasized by Immanuel Kant in his
book entitled {\it Kritik der reinen Vernunft} whose first and second
editions were published in 1781 and 1787 respectively.  However, using
his own logic, he ended up with a conclusion that there must be the
absolute inertial frame, and that we only see the frames dictated by
our subjectivity.

Einstein's special relativity was developed along Kant's line of thinking:
things depend on the frame from which you make observations.  However,
there is one big difference.  Instead of the absolute frame, Einstein
introduced an extra dimension.  Let us illustrate this using a Coca-Cola
can.  It appears like a circle if you look at it from the top, while
it appears as a rectangle from the side.  The real thing is a
three-dimensional circular cylinder.  While Kant was obsessed with
the absoluteness of the real thing, Einstein was able to observe the
importance of the extra dimension.

I was fortunate enough to be close to Eugene Wigner, and enjoyed the
privilege of asking him many questions.  I once asked him whether he
thinks like Immanuel Kant.  He said {\it Yes}.  I then asked him whether
Einstein was a Kantianist in his opinion.  Wigner said very firmly
{\it Yes}. I then asked him whether he studied the philosophy of Kant
while he was in college.  He said {\it No}, and said that he realized he
had been a Kantianist after writing so many papers in physics.  He added
that philosophers do not dictate people how to think, but their job is
to describe systematically how people think~\cite{marx}.  Wigner told me
that I was the only one who asked him this question, and asked me how I
knew the Kantian way of reasoning was working in his mind.  I gave him
the following answer.

I never had any formal education in oriental philosophy, but I know
that my frame of thinking is affected by my Korean background.  One
important aspect is that Immanuel Kant's name is known to every
high-school graduate in Korea, while he is unknown to Americans,
particularly to American physicists.  The question then is whether
there is in Eastern culture a ``natural frequency'' which can resonate
with one of the frequencies radiated from the Kantian school of
thought developed in Europe.

I would like to answer this question in the following way.  Koreans
absorbed a bulk of Chinese culture during the period of the Tang
dynasty (618-907 AD).  At that time, China was the center of the world
as the United States is today.  This dynasty's intellectual life was
based on Taoism which tells us, among others, that everything in this
universe has to be balanced between its plus (or bright) side and its
minus (or dark) side.  This way of thinking forces us to look at things
from two different or opposite directions.  This aspect of Taoism could
constitute a ``natural frequency'' which can be tuned to the Kantian
view of the world where things depend how they are observed.

I would like to point out that Hideki Yukawa was quite fond of Taoism
and studied systematically the books of Laotse and Chuangtse who were
the founding fathers of Taoism~\cite{tani79}.  Both Laotse and Chuangtse
lived before the time of Confucius.  It is interesting to note that
Kantianism is also popular is Japan, and it is my assumption that Kant's
books were translated into Japanese by Japanese philosophers first,
and Koreans of my father's age learned about Kant by reading the
translated versions.  My publication record will indicate that I studied
Yukawa's papers before becoming seriously interested in Wignerism.
I picked up a signal of possible connection between Kantianism and
Taoism while reading Yukawa's papers carefully, and this led to my
bold venture to ask Wigner whether he was a Kantianist.

Kant wrote his books in German, but he was born and spent his entire
life in a Baltic enclave now called Kaliningrad located between Poland
and Lithuania.  Historically, this place was dominated by several
different countries with different ideologies~\cite{apple94}.  However,
Kant's view was that the people there may appear differently depending
on who looks at them, but they remain unchanged.  At the same time,
they had to entertain different ideologies imposed by different rulers.
Kant translated this philosophy into physics when he discussed the
absolute and relative frames.  He was obsessed with the absolute frame,
and this is the reason why Kant is not regarded as a physicist in
Einstein's world in which we live.

The people of Kant's land stayed in the same place while experiencing
different ideological environments.  Almost like Kant, I was exposed
to two different cultural environments by moving myself from Asia to
the United States.  Thus, I often had to raise the question of
absolute and relative values.  Let us discuss this problem using one
concrete example.

About 4,500 years ago, there was a king named Yao in China.  While he
was looking for a man who could serve as the prime minister, he heard
from many people that a person named Shiyu was widely respected and
had a deep knowledge of the world.  The king then sent his messengers
to invite Shiyu to come to his palace and to run the country.  After
hearing the king's message, Shiyu without saying anything went to a
creek in front of his house and started washing his ears.  He thought
he heard the dirtiest story in his life.

Shiyu is still respected in the Eastern world as one of the wisest men
in history.  We do not know whether this person existed or is a made-up
personality.  In either case, we are led to look for a similar person
in the Western world.  In ancient Greece, each city was run by its
city council.  As we
experience even these days, people accomplish very little in committee
meetings.  Thus, it is safe to assume that the city councils in ancient
Greece did not handle matters too efficiently.  For this reason, there
was a well-respected wiseman like Shiyu who never attended his city
council meetings.  His name was Idiot.  Idiot was a wiseman, but he
never contributed his wisdom to his community.  His fellow citizens
labeled him as a useless person.  This was how the word idiot was
developed in the Western world.

We can now illustrate the difference between Lorentz invariance and
Lorentz covariance.  Idiot and Shiyu had the same personality if they
were not the same person.  The fact that they are the same person is
an illustration of the invariance.  On the other hand, they look
quite different in the worlds with two different cultural backgrounds.
This is what the covariance is about.

Idiot is a useless person in state-centered societies like Sparta.
The same person is regarded as the ultimate wiseman in a
self-centered society like Korea.  I cannot say that I know everything
about other Asian countries, but I have a deep knowledge of Korea where
I was born and raised.  The same person looks quite differently to
observers in different cultural frames.  While doing research in the
United States with my Eastern background, I was frequently forced to
find a common ground for two seemingly different views.  This cultural
background strongly influenced me in producing an expanded content of
Einstein's $E = mc^{2}$ tabulated in Table I~\cite{kim89}.

\begin{table}

\caption{Further contents of Einstein's $E = mc^{2}$.  Massive and
massless particles have different energy-momentum relations.  Einstein's
special relativity gives one relation for both.  Wigner's little group
unifies the internal space-time symmetries for massive and massless
particles which are locally isomorphic to $O(3)$ and $E(2)$ respectively.
It is a great challenge for us to find another unification.  In this
note, we present a unified picture of the quark and parton models which
are applicable to slow and ultra-fast hadrons respectively.}

\vspace{3mm}

\begin{tabular}{cccc}

{}&{}&{}&{}\\
{} & Massive, Slow \hspace*{1mm} & COVARIANCE \hspace*{1mm}&
Massless, Fast \\[4mm]\hline
{}&{}&{}&{}\\
Energy- & {}  & Einstein's & {} \\
Momentum & $E = p^{2}/2m$ & $ E = [p^{2} + m^{2}]^{1/2}$ & $E = cp$
\\[4mm]\hline
{}&{}&{}&{}\\
Internal & $S_{3}$ & {}  &  $S_{3}$ \\[-1mm]
space-time &{} & Wigner's  & {} \\ [-1mm]
symmetry & $S_{1}, S_{2}$ & Little Group & Gauge
Transformations \\[4mm]\hline
{}&{}&{}&{}\\
Relativistic & {} & {} & {} \\[-1mm]
Extended & Quark Model & Covariant Model of Hadrons & Partons \\ [-1mm]
Particles & {} & {} & {} \\[2mm]
\end{tabular}
\end{table}

Let us go back to the question of relative values.  For Taoists, those
two opposite faces of the same person is like ``yang'' (plus) versus
``ying'' (minus).  Finding the harmony between these two opposite
points of view is the ideal way to live in this world.  We cannot
always live like Shiyu, nor like Idiot.  The key to happiness is to
find a harmony between the individual and the society to which
he/she belongs.  The key word here seems to be ``harmony.''

To Kantianists, however, it is quite natural for the same character
to appear differently in two different environments.  The problem is
to find the absolute value from these two different faces.  Does this
absolute value exist?  According to Kant, it exists.  To most of us,
it is very difficult to find it if it exists.

Let us finally visit Einstein.  He avoids the question of the existence
of the absolute value.  Instead, he introduces a new variable.  The
variable is the ratio between the individual's ability to contribute
and the community's need for his service.  The best way to live in this
world is to adjust this variable to the optimal value.  Einstein's
approach is to a quantification of Taoism by introducing a new
variable.

If Taoism is so close to Einsteinism, why do we have to mention Kant
at all?  We have to keep in mind that Kant was the first person who
formulated the idea that observers can participate in drawing the
picture of the world.  It is not clear whether Einstein could have
formulated his relativity theory without Kant.  Kant spent many
years for studying physics, namely observer-dependent physics.  However,
because of his obsession toward the absolute thing, he spent all of
his time for finding the absolute frame.  If one has a Taoist background,
he/she is more likely to appreciate the concept of relativistic
covariance.

I would like to stress that Taoism is not confined to the ancient
Eastern world.  It is practiced frequently in the United States.
Let us look at American football games.  The offensive strategy does
not rely on brute force, but is aimed at breaking the harmony of the
defense.  For instance, when the offensive team is near the end zone,
the defense becomes very strong because it covers only a small area.
Then, it is not uncommon for the offense to place four wide-receivers
instead of two.  This will divide the defense into two sides while
creating a hole in the middle.  Then the quarter-back can carry the
ball to the end zone.  The key word is to destroy the balance of the
defense.

Taoism forms the philosophical base for Sun Tzu's classic book on
military arts~\cite{suntzu96}.  When I watch the football games, I
watch them as Sun-Tzu games.  My maternal grandfather was fluent in
the Chinese classic literature, and he was particularly fond of Sun Tzu.
He told me many stories from Sun Tzu's books.  This presumably was how
I inherited some of the Taoist tradition.  Needless to say, my research
life was influenced by my Asian background.  Many of my Asian friends
complain that they are handicapped to do original research because of
the East-West cultural difference.  I disagree with them.  This
difference could be the richest source of originality.

\section{Little Groups of the Poincar\'e Group}\label{littleg}

The Poincar\'e group is the group of inhomogeneous Lorentz
transformations, namely Lorentz transformations preceded or followed
by space-time translations.  In order to study this group, we have to
understand first the group of Lorentz transformations, the group of
translations, and how these two groups are combined to form the
Poincar\'e group.  The Poincar\'e group is a semi-direct product of
the Lorentz and translation groups.  The two Casimir operators of
this group correspond to the (mass)$^{2}$ and (spin)$^{2}$ of a given
particle.  The particle mass and its spin magnitude are
Lorentz-invariant quantities.  Then what are the Lorentz-covariant
entities.

The question is how to construct the representations of the Lorentz
group which are relevant to physics.  With this point in mind, Wigner
in 1939 studied the subgroups of the Lorentz group whose
transformations leave the four-momentum of a given free
particle~\cite{wig39}.  The maximal subgroup of the Lorentz group
which leaves the four-momentum invariant is called the little group.
Since the little group leaves the four-momentum invariant, it governs
the internal space-time symmetries of relativistic particles.  Wigner
shows in his paper that the internal space-time symmetries of massive
and massless particles are dictated by the $O(3)$-like and $E(2)$-like
little groups respectively.

The $O(3)$-like little group is locally isomorphic to the
three-dimensional rotation group, which is very familiar to us.
For instance, the group $SU(2)$ for the electron spin is an
$O(3)$-like little group.  The group $E(2)$ is the Euclidean group in
a two-dimensional space, consisting of translations and rotations on a
flat surface.  We are performing these transformations everyday on
ourselves when we move from home to school.  The mathematics of these
Euclidean transformations are also simple.  However, the group of these
transformations are not well known to us.  In Appendix~\ref{o3e2}, we
give a matrix representation of the $E(2)$ group.

The group of Lorentz transformations consists of three boosts and
three rotations.  The rotations therefore constitute a subgroup of
the Lorentz group.  If a massive particle is at rest, its four-momentum
is invariant under rotations.  Thus the little group for a massive
particle at rest is the three-dimensional rotation group.  Then what is
affected by the rotation?  The answer to this question is very simple.
The particle in general has its spin.  The spin orientation is going
to be affected by the rotation!

If the rest-particle is boosted along the $z$ direction, it will pick
up a non-zero momentum component.  The generators of the $O(3)$ group
will then be boosted.  The boost will take the form of conjugation by
the boost operator.  This boost will not change the Lie algebra of the
rotation group, and the boosted little group will still leave the
boosted four-momentum invariant.  We call this the $O(3)$-like little
group.  If we use the four-vector coordinate $(x, y, z, t)$, the
four-momentum vector for the particle at rest is $(0, 0, 0, m)$, and
the three-dimensional rotation group leaves this four-momentum invariant.
This little group is generated by
\begin{equation}
J_{1} = \pmatrix{0&0&0&0\cr0&0&-i&0\cr0&i&0&0\cr0&0&0&0} , \qquad
J_{2} = \pmatrix{0&0&i&0\cr0&0&0&0\cr-i&0&0&0\cr0&0&0&0} ,
\end{equation}
and
\begin{equation}\label{j3}
J_{3} = \pmatrix{0 & -i & 0 & 0 \cr i & 0 & 0 & 0
\cr 0 & 0 & 0 & 0 \cr 0 & 0 & 0 & 0} ,
\end{equation}
which satisfy the commutation relations:
\begin{equation}
[J_{i}, J_{j}] = i\epsilon_{ijk} J_{k} .
\end{equation}

It is not possible to bring a massless particle to its rest frame.
In his 1939 paper~\cite{wig39}, Wigner observed that the little group
for a massless particle moving along the $z$ axis is generated by the
rotation generator around the $z$ axis, namely $J_{3}$ of Eq.(\ref{j3}),
and two other generators which take the form
\begin{equation}\label{n1n2}
N_{1} = \pmatrix{0 & 0 & -i & i \cr 0 & 0 & 0 & 0
\cr i & 0 & 0 & 0 \cr i & 0 & 0 & 0} ,  \quad
N_{2} = \pmatrix{0 & 0 & 0 & 0 \cr 0 & 0 & -i & i
\cr 0 & i & 0 & 0 \cr 0 & i & 0 & 0} .
\end{equation}
If we use $K_{i}$ for the boost generator along the i-th axis, these
matrices can be written as
\begin{equation}
N_{1} = K_{1} - J_{2} , \qquad N_{2} = K_{2} + J_{1} ,
\end{equation}
with
\begin{equation}
K_{1} = \pmatrix{0&0&0&i\cr0&0&0&0\cr0&0&0&0\cr i&0&0&0} , \qquad
K_{2} = \pmatrix{0&0&0&0\cr0&0&0&i\cr0&0&0&0\cr0&i&0&0} .
\end{equation}
The generators $J_{3}, N_{1}$ and $N_{2}$ satisfy the following set
of commutation relations.
\begin{equation}\label{e2lcom}
[N_{1}, N_{2}] = 0 , \quad [J_{3}, N_{1}] = iN_{2} ,
\quad [J_{3}, N_{2}] = -iN_{1} .
\end{equation}
In Appendix \ref{o3e2}, we discuss the generators of the $E(2)$ group.
They are $J_{3}$ which generates rotations around the $z$ axis, and
$P_{1}$ and $P_{2}$ which generate translations along the $x$ and $y$
directions respectively.  If we replace $N_{1}$ and $N_{2}$ by $P_{1}$
and $P_{2}$, the above set of commutation relations becomes the set
given for the $E(2)$ group given in Eq.(\ref{e2com}).  This is the
reason why we say the little group for massless particles is
$E(2)$-like.  Very clearly, the matrices $N_{1}$ and $N_{2}$ generate
Lorentz transformations.

It is not difficult to associate the rotation generator $J_{3}$ with
the helicity degree of freedom of the massless particle.   Then what
physical variable is associated with the $N_{1}$ and $N_{2}$
generators?  Wigner was the one who discovered the existence of these
generators, but did not give any physical interpretation to these
translation-like generators.  For this reason, for many years, only
those representations with the zero-eigenvalues of the $N$ operators
were thought to be physically meaningful representations~\cite{wein64}.
It was not until 1971 when Janner and Janssen reported that the
transformations generated by these operators are gauge
transformations~\cite{janner71,kim97poz}.  The role of this
translation-like transformation has also been studied for spin-1/2
particles, and it was concluded that the polarization of neutrinos
is due to gauge invariance~\cite{hks82,kim97min}.

Another important development along this line of research is the
application of group contractions to the unifications of the two
different little groups for massive and massless particles.
We always associate the three-dimensional rotation group with a spherical
surface.  Let us consider a circular area of radius 1 kilometer centered
on the north pole of the earth.  Since the radius of the earth is more
than 6,450 times longer, the circular region appears flat.  Thus, within
this region, we use the $E(2)$ symmetry group for this region.  The
validity of this approximation depends on the ratio of the two radii.

In 1953, Inonu and Wigner formulated this problem as the contraction of
$O(3)$ to $E(2)$~\cite{inonu53}.  How about then the little groups which
are isomorphic to $O(3)$ and $E(2)$?  It is reasonable to expect that the
$E(2)$-like little group be obtained as a limiting case for of the
$O(3)$-like little group for massless particles.  In 1981, it was
observed by Ferrara and Savoy that this limiting process is the Lorentz
boost \cite{ferrara82}.  In 1983, using the
same limiting process as that of Ferrara and Savoy, Han {\it et al.}
showed that transverse rotation generators become the generators of
gauge transformations in the limit of infinite momentum and/or zero mass
\cite{hks83pl}.  In 1987, Kim and Wigner showed that the little group for
massless particles is the cylindrical group which is isomorphic to the
$E(2)$ group~\cite{kiwi87jm}.  This completes the second row in Table I,
where Wigner's little group unifies the internal space-time symmetries
of massive and massless particles.

We are now interested in constructing the third row in Table I.  As we
promised in Sec.~\ref{intro}, we will be dealing with hadrons which are
bound states of quarks with space-time extensions.  For this purpose, we
need a set of covariant wave functions consistent with the existing laws
of quantum mechanics, including of course the uncertainty principle and
probability interpretation.

With these wave functions, we propose to solve the following problem in
high-energy physics.  The quark model works well when hadrons are at
rest or move slowly.  However, when they move with speed close to that
of light, they appear as a collection of infinite-number of
partons~\cite{fey69}.  As we stated above, we need a set of wave
functions which can be Lorentz-boosted.  How can we then construct such
a set?  In constructing wave functions for any purpose in quantum
mechanics, the standard procedure is to try first harmonic oscillator
wave functions.  In studying the Lorentz boost, the standard language
is the Lorentz group.  Thus the first step to construct covariant wave
functions is to work out representations of the Lorentz group using
harmonic oscillators~\cite{dir45,knp86}.

\section{Covariant Harmonic Oscillators}\label{covham}

If we construct a representation of the Lorentz group using normalizable
harmonic oscillator wave functions, the result is the covariant harmonic
oscillator formalism~\cite{knp86}.  The formalism constitutes a
representation of Wigner's $O(3)$-like little group for a massive
particle with internal space-time structure.  This oscillator formalism
has been shown to be effective in explaining the basic phenomenological
features of relativistic extended hadrons observed in high-energy
laboratories.  In particular, the formalism shows that the quark model
and Feynman's parton picture are two different manifestations of one
covariant entity~\cite{knp86,kim89}.  The essential feature of the
covariant harmonic oscillator formalism is that Lorentz boosts are
squeeze transformations~\cite{kn73,knp91}.  In the light-cone coordinate
system, the boost transformation expands one coordinate while contracting
the other so as to preserve the product of these two coordinate remains
constant.  We shall show that the parton picture emerges from this
squeeze effect.

Let us consider a bound state of two particles.  For convenience, we
shall call the bound state the hadron, and call its constituents quarks.
Then there is a Bohr-like radius measuring the space-like separation
between the quarks.  There is also a time-like separation between the
quarks, and this variable becomes mixed with the longitudinal spatial
separation as the hadron moves with a relativistic speed.  There are
no quantum excitations along the time-like direction.  On the other
hand, there is the time-energy uncertainty relation which allows
quantum transitions.  It is possible to accommodate these aspect within
the framework of the present form of quantum mechanics.  The uncertainty
relation between the time and energy variables is the c-number
relation~\cite{dir27},
which does not allow excitations along the time-like coordinate.  We
shall see that the covariant harmonic oscillator formalism accommodates
this narrow window in the present form of quantum mechanics.

For a hadron consisting of two quarks, we can consider their space-time
positions $x_{a}$ and $x_{b}$, and use the variables
\begin{equation}
X = (x_{a} + x_{b})/2 , \qquad x = (x_{a} - x_{b})/2\sqrt{2} .
\end{equation}
The four-vector $X$ specifies where the hadron is located in space and
time, while the variable $x$ measures the space-time separation between
the quarks.  In the convention of Feynman {\it et al.}~\cite{fkr71},
the internal motion of the quarks bound by a harmonic oscillator
potential of unit strength can be described by the Lorentz-invariant
equation
\begin{equation}\label{osceq}
{1\over 2}\left\{x^{2}_{\mu} -
{\partial ^{2} \over \partial x_{\mu}^{2}}
\right\} \psi(x)= \lambda \psi(x) .
\end{equation}
It is now possible to construct a representation of the Poincar\'e group
from the solutions of the above differential equation~\cite{knp86}.

The coordinate $X$ is associated with the overall hadronic
four-momentum, and the space-time separation variable $x$ dictates
the internal space-time symmetry or the $O(3)$-like little group.  Thus,
we should construct the representation of the little group from the
solutions of the differential equation in Eq.(\ref{osceq}).  If the
hadron is at rest, we can separate the $t$ variable from the equation.
For this variable we can assign the ground-state wave function to
accommodate the c-number time-energy uncertainty relation~\cite{dir27}.
For the three space-like variables, we can solve the oscillator
equation in the spherical coordinate system with usual orbital and
radial excitations.  This will indeed constitute a representation of
the $O(3)$-like little group for each value of the mass.  The solution
should take the form
\begin{equation}
\psi(x,y,z,t) = \psi(x,y,z) \left({1\over \pi}\right)^{1/4}
\exp\left(-t^{2}/2 \right) ,
\end{equation}
where $\psi(x,y,z)$ is the wave function for the three-dimensional
oscillator with appropriate angular momentum quantum numbers.
The above wave function constitutes a representation of Wigner's
$O(3)$-like little group for a massive particle~\cite{knp86}.

Since the three-dimensional oscillator differential equation is
separable in both spherical and Cartesian coordinate systems,
$\psi(x,y,z)$ consists of Hermite polynomials of $x, y$, and $z$.
If the Lorentz boost is made along the $z$ direction, the $x$ and $y$
coordinates are not affected, and can be temporarily dropped from the wave
function.  The wave function of interest can be written as
\begin{equation}
\psi^{n}(z,t) = \left({1\over\pi}\right)^{1/4}\exp\left(-t^{2}/2\right)
\phi_{n}(z) ,
\end{equation}
where $\phi_{n}(z)$ is the one-dimensional oscillator wave function for
the n-th excited states, which takes the form
\begin{equation}\label{phi}
\phi_{n}(z) = \left({1\over\pi n!2^{n}}\right)^{1/2}H_{n}(z)
\exp (-z^{2}/2) .
\end{equation}
The full wave function $\psi ^{n}(z,t)$ is
\begin{equation}\label{fullwf}
\psi^{n}_{0}(z,t) = \left({1\over \pi n! 2^{n}}\right)^{1/2} H_{n}(z)
\exp \left\{-{1\over 2}\left(z^{2} + t^{2} \right) \right\} .
\end{equation}
The subscript $0$ means that the wave function is for the hadron at rest.
The above expression is not Lorentz-invariant, and its localization
undergoes a Lorentz squeeze as the hadron moves along the $z$
direction~\cite{knp86}.

It is convenient to use the light-cone variables to describe Lorentz
boosts.  The light-cone coordinate variables are
\begin{equation}
u = (z + t)/\sqrt{2} , \qquad v = (z - t)/\sqrt{2} .
\end{equation}
In terms of these variables, the Lorentz boost along the $z$
direction,
\begin{equation}
\pmatrix{z' \cr t'} = \pmatrix{\cosh \eta & \sinh \eta \cr
\sinh \eta & \cosh \eta} \pmatrix{z \cr t} ,
\end{equation}
takes the simple form
\begin{equation}\label{lorensq}
u' = e^{\eta} u , \qquad v' = e^{-\eta} v ,
\end{equation}
where $\eta $ is the boost parameter and is $\tanh ^{-1}(v/c)$.
The $u$ variable becomes expanded while the $v$ variable
becomes contracted.  This is the squeeze mechanism illustrated
discussed extensively in the literature~\cite{kn73,knp91}.  This
squeeze transformation is also illustrated in Fig.~1.

The wave function of Eq.(\ref{fullwf}) can be written as
\begin{equation}\label{wf10}
\psi^{n}_{0}(z,t) = \left({1 \over \pi n!2^{n}} \right)^{1/2}
H_{n}\left((u + v)/\sqrt{2}\right)
\exp\left\{-{1\over 2} (u^{2} + v^{2}) \right\} .
\end{equation}
If the system is boosted, the wave function becomes
\begin{equation}\label{wf11}
\psi^{n}_{\eta}(z,t) = \left({1 \over \pi n!2^{n}} \right)^{1/2}
H_{n} \left((e^{-\eta}u + e^{\eta}v)/\sqrt{2} \right)
\times \exp \left\{-{1\over 2}\left(e^{-2\eta}u^{2} +
e^{2\eta}v^{2}\right)\right\} .
\end{equation}
In both Eqs. (\ref{wf10}) and (\ref{wf11}), the localization
property of the wave function in the $u v$ plane is determined by
the Gaussian factor, and it is sufficient to study the ground state
only for the essential feature of the boundary condition.  The wave
functions in Eq.(\ref{wf10}) and Eq.(\ref{wf11}) then respectively
become
\begin{equation}\label{wf13}
\psi_{0}(z,t) = \left({1 \over \pi} \right)^{1/2}
\exp \left\{-{1\over 2} (u^{2} + v^{2}) \right\} .
\end{equation}
If the system is boosted, the wave function becomes
\begin{equation}\label{wf14}
\psi_{\eta}(z,t) = \left({1 \over \pi}\right)^{1/2}
\exp \left\{-{1\over 2}\left(e^{-2\eta}u^{2} +
e^{2\eta}v^{2}\right)\right\} .
\end{equation}
We note here that the transition from Eq.(\ref{wf13}) to Eq.(\ref{wf14}) is a
squeeze transformation.  The wave function of Eq.(\ref{wf13}) is distributed
within a circular region in the $u v$ plane, and thus in the $z t$ plane.
On the other hand, the wave function of Eq.(\ref{wf14}) is distributed in an
elliptic region.  This ellipse is a ``squeezed'' circle with the same area
as the circle, as is illustrated in Fig.~1.

\begin{figure}[thb]  
\centerline{\psfig{figure=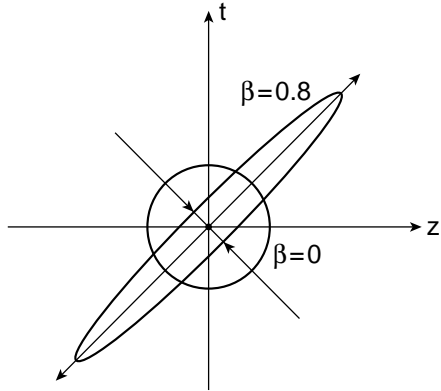,angle=0,height=60mm}}
\caption{Effect of the Lorentz boost on the space-time wave function.
The circular space-time distribution at the rest frame becomes
Lorentz-squeezed to become an elliptic distribution.}
\end{figure}


\section{Feynman's Parton Picture}\label{parton}

It is safe to believe that hadrons are quantum bound states of quarks
having localized probability distribution.  As in all bound-state cases,
this localization condition is responsible for the existence of
discrete mass spectra.  The most convincing evidence for this
bound-state picture is the hadronic mass spectra which are observed in
high-energy laboratories~\cite{fkr71,knp86}.
However, this picture of bound states is applicable only to observers
in the Lorentz frame in which the hadron is at rest.  How would the
hadrons appear to observers in other Lorentz frames?  More specifically,
can we use the picture of Lorentz-squeezed hadrons discussed in
Sec.~\ref{covham}.

Proton's radius is $10^{-5}$ of that of the hydrogen atom. Therefore,
it is not unnatural to assume that the proton has a point charge in
atomic physics.  However, while carrying out experiments on electron
scattering from proton targets, Hofstadter in 1955 observed that the
proton charge is spread out~\cite{hofsta55}.  In this experiment,
an electron emits
a virtual photon, which then interacts with the proton.  If the proton
consists of quarks distributed within a finite space-time region,
the virtual photon will interact with quarks which carry fractional
charges.  The scattering amplitude will depend on the way in which
quarks are distributed within the proton.  The portion of the scattering
amplitude which describes the interaction between the virtual photon
and the proton is called the form factor.

Although there have been many attempts to explain this phenomenon
within the framework of quantum field theory, it is quite natural
to expect that the wave function in the quark model will describe
the charge distribution.  In high-energy experiments, we are dealing
with the situation in which the momentum transfer in the scattering
process is large.  Indeed, the Lorentz-squeezed wave functions lead
to the correct behavior of the hadronic form factor for large values
of the momentum transfer~\cite{fuji70}.

While the form factor is the quantity which can be extracted from the
elastic scattering, it is important to realize that in high-energy
processes, many particles are produced in the final state.  They are called
inelastic processes.  While the elastic process is described by the total
energy and momentum transfer in the center-of-mass coordinate system, there
is, in addition, the energy transfer in inelastic scattering.  Therefore, we
would expect that the scattering cross section would depend on the energy,
momentum transfer, and energy transfer.  However, one prominent feature in
inelastic scattering is that the cross section remains nearly constant for a
fixed value of the momentum-transfer/energy-transfer ratio.  This
phenomenon is called ``scaling''~\cite{bj69}.

\begin{figure}[thb]  
\centerline{\psfig{figure=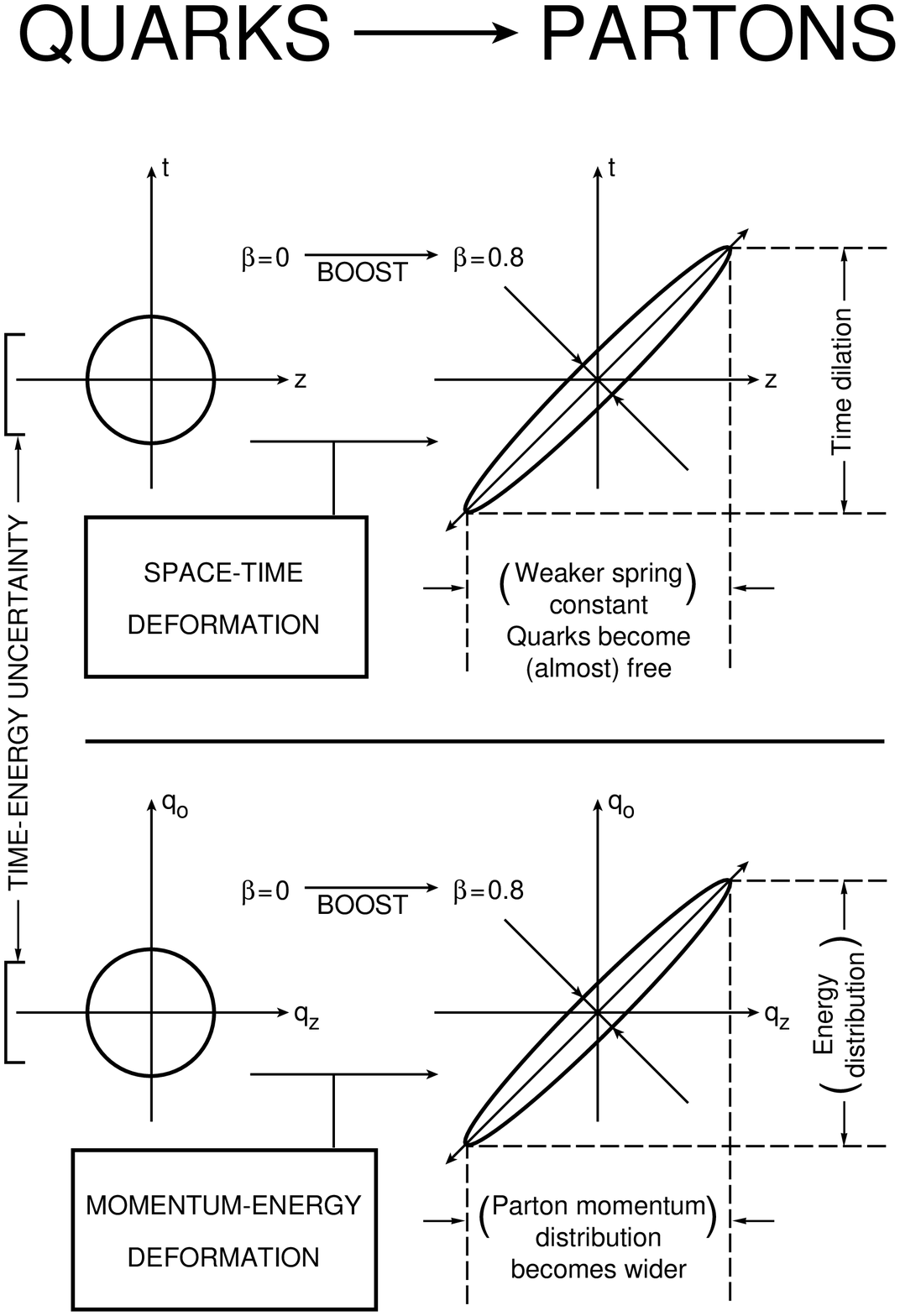,angle=0,height=140mm}}
\vspace{5mm}
\caption{Lorentz-squeezed space-time and momentum-energy wave functions.
As the hadron's speed approaches that of light, both wave functions
become concentrated along their respective positive light-cone axes.
These light-cone concentrations lead to Feynman's parton picture.}
\end{figure}

In order to explain the scaling behavior in inelastic scattering, Feynman in
1969 observed that a fast-moving hadron can be regarded as a collection of
many ``partons'' whose properties do not appear to be identical to those of
quarks~\cite{fey69}.  For example, the number of quarks inside a static
proton is three, while the number of partons in a rapidly moving proton
appears to be infinite.  The question then is how the proton looking like a
bound state of quarks to one observer can appear different to an observer in
a different Lorentz frame?  Feynman made the following systematic
observations.

    a). The picture is valid only for hadrons moving with velocity close
       to that of light.

    b). The interaction time between the quarks becomes dilated, and
        partons\\
\hspace{20mm} behave as free independent particles.

    c). The momentum distribution of partons becomes widespread as the
       hadron\\
\hspace{20mm} moves fast.

    d). The number of partons seems to be infinite or much larger than that
       of quarks.

\noindent Because the hadron is believed to be a bound state of two or three
quarks, each of the above phenomena appears as a paradox, particularly b) and
c) together.  We would like to resolve this paradox using the covariant
harmonic oscillator formalism.

For this purpose, we need a momentum-energy wave function.  If the quarks
have the four-momenta $p_{a}$ and $p_{b}$, we can construct two independent
four-momentum variables~\cite{fkr71}
\begin{equation}
P = p_{a} + p_{b} , \qquad q = \sqrt{2}(p_{a} - p_{b}) .
\end{equation}
The four-momentum $P$ is the total four-momentum and is thus the hadronic
four-momentum.  $q$ measures the four-momentum separation between the quarks.

We expect to get the momentum-energy wave function by taking the Fourier
transformation of Eq.(\ref{wf14}):
\begin{equation}\label{fourier}
\chi_{\eta}(q_{z},q_{0}) = \left({1 \over 2\pi}\right)
\int \psi_{\eta}(z,t) \exp{\left\{-i(q_{z}z - q_{0}t)\right\}} dx dt .
\end{equation}
Let us now define the momentum-energy variables in the light-cone coordinate
system as
\begin{equation}\label{conju}
q_{u} = (q_{0} - q_{z})/\sqrt{2} ,  \qquad
q_{v} = (q_{0} + q_{z})/\sqrt{2} .
\end{equation}
In terms of these variables, the Fourier transformation of
Eq.(\ref{fourier}) can be written as
\begin{equation}\label{fourier2}
\chi_{\eta}(q_{z},q_{0}) = \left({1 \over 2\pi}\right)
\int \psi_{\eta}(z, t) \exp{\left\{-i(q_{u} u + q_{v} v)\right\}} du dv .
\end{equation}
The resulting momentum-energy wave function is
\begin{equation}
\chi_{\eta}(q_{z},q_{0}) = \left({1 \over \pi}\right)^{1/2}
\exp\left\{-{1\over 2}\left(e^{-2\eta}q_{u}^{2} +
e^{2\eta}q_{v}^{2}\right)\right\} .
\end{equation}
Because we are using here the harmonic oscillator, the mathematical form
of the above momentum-energy wave function is identical to that of the
space-time wave function.  The Lorentz squeeze properties of these wave
functions are also the same, as are indicated in Fig.~2.

When the hadron is at rest with $\eta = 0$, both wave functions behave like
those for the static bound state of quarks.  As $\eta$ increases, the wave
functions become continuously squeezed until they become concentrated along
their respective positive light-cone axes.  Let us look at the z-axis
projection of the space-time wave function.  The width of the quark
distribution increases as the hadronic speed approaches that of the speed of
light.  The position of each quark appears widespread to the observer in the
laboratory frame, and the quarks appear like free particles.

Furthermore, interaction time of the quarks among themselves become dilated.
Because the wave function becomes wide-spread, the distance between one end
of the harmonic oscillator well and the other end increases as is indicated
in Fig.~2.  This effect, first noted by Feynman~\cite{fey69}, is universally
observed in high-energy hadronic experiments.  The period is oscillation is
increases like $e^{\eta}$.  On the other hand, the interaction time with
the external signal, since it is moving in the direction opposite to the
direction of the hadron, it travels along the negative light-cone axis.  If
the hadron contracts along the negative light-cone axis, the interaction time
decreases by $e^{-\eta}$.  The ratio of the interaction time to the
oscillator period becomes $e^{-2\eta}$.  The energy of each proton coming
out of the Fermilab accelerator is $900 GeV$.  This leads the ratio to
$10^{-6}$.  This is indeed a small number.  The external signal is not able
to sense the interaction of the quarks among themselves inside the hadron.

The momentum-energy wave function is just like the space-time wave function.
The longitudinal momentum distribution becomes wide-spread as the hadronic
speed approaches the velocity of light.  This is in contradiction with our
expectation from nonrelativistic quantum mechanics that the width of the
momentum distribution is inversely proportional to that of the position wave
function.  Our expectation is that if the quarks are free, they must have
their sharply defined momenta, not a wide-spread distribution.  This apparent
contradiction presents to us the following two fundamental questions:

\begin{itemize}

   \item[a].  If both the spatial and momentum distributions become
      widespread as the hadron moves, and if we insist on Heisenberg's
      uncertainty relation, is Planck's constant dependent on the
      hadronic velocity?

   \item[b].  Is this apparent contradiction related to another apparent
      contradiction that the number of partons is infinite while there
      are only two or three quarks inside the hadron?
\end{itemize}
The answer to the first question is {\it No}, and that for the second
question is {\it Yes}.  Let us answer the first question which is
related to the Lorentz invariance of Planck's constant.  If we take the
product of the width of the longitudinal momentum distribution and that
of the spatial distribution, we end up with the relation
\begin{equation}
<z^{2}><q_{z}^{2}> = (1/4)[\cosh(2\eta)]^{2}  .
\end{equation}
The right-hand side increases as the velocity parameter increases.  This
could lead us to an erroneous conclusion that Planck's constant becomes
dependent on velocity.  This is not correct, because the longitudinal
momentum variable $q_{z}$ is no longer conjugate to the longitudinal
position variable when the hadron moves.

In order to maintain the Lorentz-invariance of the uncertainty product,
we have to work with a conjugate pair of variables whose product does
not depend on the velocity parameter.  Let us go back to Eq.(\ref{conju})
and Eq.(\ref{fourier2}).  It is quite clear that the light-cone variable
$u$ and $v$ are conjugate to $q_{u}$ and $q_{v}$ respectively.  It is
also clear that the distribution along the $q_{u}$ axis shrinks as the
$u$-axis distribution expands.  The exact calculation leads to
\begin{equation}
<u^{2}><q_{u}^{2}> = 1/4 , \qquad  <v^{2}><q_{v}^{2}> = 1/4  .
\end{equation}
Planck's constant is indeed Lorentz-invariant.

Let us next resolve the puzzle of why the number of partons appears to
be infinite while there are only a finite number of quarks inside the
hadron.  As the hadronic speed approaches the speed of light, both the
x and q distributions become concentrated along the positive light-cone
axis.  This means that the quarks also move with velocity very close
to that of light.  Quarks in this case behave like massless particles.

We then know from statistical mechanics that the number of massless
particles is not a conserved quantity.  For instance, in black-body
radiation, free light-like particles have a widespread momentum
distribution.  However, this does not contradict the known principles
of quantum mechanics, because the massless photons can be divided into
infinitely many massless particles with a continuous momentum
distribution.

Likewise, in the parton picture, massless free quarks have a wide-spread
momentum distribution.  They can appear as a distribution of an
infinite number of free particles.  These free massless particles are the
partons.  It is possible to measure this distribution in high-energy
laboratories, and it is also possible to calculate it using the covariant
harmonic oscillator formalism.  We are thus forced to compare these two
results.  According to Hussar's calculation~\cite{hussar81},
the Lorentz-boosted oscillator wave function produces a reasonably
accurate parton distribution.

\section{Feynman's Rest of the Universe}\label{rest}

We have seen in Sec. \ref{parton} that the time-separation variable
plays the pivotal role in the light-cone formulation and Feynman's
parton picture.  Then, do measure this variable or distribution along
this variable?  The answer to this question is clearly {\it No}.  Does
the present form of quantum mechanics accommodate this non-measurable
variable?  The answer is {\it Yes}, and is an increase in entropy.

The entropy is a measure of our ignorance and is computed from the
density matrix~\cite{neu32,wiya63}.  The density matrix is needed
when the experimental procedure does not analyze all relevant variables
to the maximum extent consistent with quantum mechanics~\cite{fano57}.
The purpose of the present note is to discuss a concrete example of the
entropy arising from our ignorance in relativistic quantum mechanics.

As was discussed in the literature for several different purposes,
this wave function can be expanded as~\cite{knp86}
\begin{equation}
\psi_{\eta}(z,t) = (1/\cosh\eta)\sum^{}_{k}
(\tanh\eta)^{k} \phi_{k}(z) \phi_{k}^{*}(t) ,
\end{equation}
where $\phi_{k}(z)$ is the $k$-th excited oscillator wave function
given in Eq.(\ref{phi}).  From this expression, we can construct the
pure-state density matrix
\begin{equation}\label{pure}
\rho_{\eta}(z,t;z',t') = \psi_{\eta}(z,t)\psi_{\eta}^{*}(z',t') ,
\end{equation}
which satisfies the condition $\rho^{2} = \rho $:
\begin{equation}
\rho^{n}_{\eta}(z,t;z',t') = \int \rho^{n}_{\eta} (z,t;z'',t'')
\rho^{n}_{\eta}(z'',t'';z',t') dz'' dt'' .
\end{equation}
However, there are at present no measurement theories which accommodate
the time-separation variable t.  Thus, we can take the trace of the
$\rho$ matrix with respect to the $t$ variable.  Then the resulting
density matrix is
\begin{equation}\label{densi}
\rho_{\eta}(z,z') = \int \psi_{\eta}(z,t) \psi_{\eta}^{*}(z',t) dt
= (1/\cosh\eta)^{2}\sum^{}_{k}
(\tanh\eta)^{2k} \phi_{k}(z)\phi_{k}^{*}(z') .
\end{equation}
The trace of this density matrix is one, but the trace of $\rho^{2}$ is
less than one, as
\begin{equation}
Tr\left(\rho^{2}\right) = \int \rho_{\eta}(z,z')
\rho_{\eta}(z',z) dz'dz
= (1/\cosh\eta)^{4} \sum^{}_{k}(\tanh \eta)^{4k} ,
\end{equation}
which is less than one.  This is due to the fact that we do not
know how to deal with the time-like separation in the present formulation of
quantum mechanics.  Our knowledge is less than complete.

The standard way to measure this ignorance is to calculate the
entropy defined as~\cite{neu32,wiya63}
\begin{equation}
S = - Tr\left(\rho~\ln(\rho) \right) .
\end{equation}
\noindent If we pretend to know the distribution along the time-like
direction and use the pure-state density matrix given in Eq.(\ref{pure}),
then the entropy is zero.  However, if we do not know how to deal with
the distribution along $t$, then we should use the density matrix of
Eq.(\ref{densi}) to calculate the entropy, and the result is
\begin{equation}
S = 2\left\{(\cosh\eta)^{2}\ln(\cosh\eta) -
(\sinh\eta)^{2}\ln(\sinh\eta)\right\} .
\end{equation}
Let us go back to the wave function given in Eq.(\ref{fullwf}).
From the wave function, we can derive the density matrix by performing
the integral of Eq.(\ref{densi}).  The result is
\begin{equation}
\rho(z,z') = \left({1\over\pi\cosh 2\eta}\right)^{1/2}
\exp\left\{-{1\over 4}[(z + z')^{2}/\cosh 2\eta
+ (z - z')^{2}\cosh 2\eta] \right\} .
\end{equation}
The diagonal elements of the above density matrix is
\begin{equation}\label{diag}
\rho(z,z') = \left({1\over \pi \cosh 2\eta} \right)^{1/2}
\exp\left({-z^{2} \over \cosh 2\eta} \right) .
\end{equation}
The width of the distribution becomes $(\cosh \eta )^{1/2}$, and
becomes wide-spread as the hadronic speed increases.  Likewise, the
momentum distribution becomes wide-spread~\cite{knp86,hkn90pl}.
This simultaneous increase in the momentum and position distribution
widths is called the parton phenomenon in high-energy
physics~\cite{fey69}.  The position-momentum uncertainty
becomes $\cosh\eta$.  This increase in uncertainty is due to our
ignorance about the physical but unmeasurable time-separation variable.

The use of an unmeasurable variable as a ``shadow'' coordinate is not
new in physics and is of current interest~\cite{ume82}.  Feynman's book
on statistical mechanics contains the following paragraph~\cite{fey72}.

{\it When we solve a quantum-mechanical problem, what we really do is
divide the universe into two parts - the system in which we are
interested and the rest of the universe.  We then usually act as if the
system in which we are interested comprised the entire universe.  To
motivate the use of density matrices, let us see what happens when we
include the part of the universe outside the system.}

In this section, we have identified Feynman's rest of the universe as
the time-separation coordinate in a relativistic two-body problem.  Our
ignorance about this coordinate leads to a density matrix for a
non-pure state, and consequently to an increase of entropy.

\section*{Concluding Remarks}
The phenomenological aspects of the covariant oscillator formalism
have been extensively discussed in the literature~\cite{knp86}.  In
this paper, we used this formalism to illustrate the truth that the
Lorentz covariance is quite different from the Lorentz invariance.
The covariance we discussed in this paper is powerful enough to
resolve the question of whether quarks are partons.

The oscillator formalism constitutes a representation of Wigner's
little group governing the internal space-time symmetries of
relativistic particles.  For this purpose, we have given a
comprehensive review of the little groups for massive and massless
particles.  We have discussed also the contraction procedure in which
the $E(2)$-like little group for massless particles is obtained from
the $O(3)$-like little group for massive particles.  In so doing, we
have explained the contents of Table I.

In addition, we discussed the fundamental issue of the time-separation
variable.  It does not appear to be a significant variable in
non-relativistic quantum mechanics.  However, it plays the central role
in covariant quantum mechanics.  If we do not make observations on this
variable, it constitutes Feynman's rest of the universe.  The failure
to make measurements in this time-like direction results in an increase
in entropy.

The harmonic oscillator is the natural language for physicists.  If
combined with the Lorentz group, it can provide a powerful theoretical
framework in many different branches of physics.  One noteworthy
development has been that the oscillator representation of the Lorentz
group forms the basis for the theory of coherent and squeezed states of
light~\cite{knp91}.   It has also been shown that the six-parameter
Lorentz group constitutes the basic language for polarization
optics~\cite{hkn97josa}.  It has been shown that the Jones vector and
the Stokes parameters can be regarded as the two-component $SL(2,c)$
spinor and the Minkowskian four-vector respectively.

As is seen in Fig.~1, the Lorentz boost is a squeeze transformation.
This geometric property is not yet widely known among physicists, but
may play many important roles in the future.  For instance, most of
the soluble models in physics is based on diagonalization of coupled
oscillators, including the Lee model in quantum field theory,
the Bogoliubov transformation in superconductivity, and the covariant
oscillator model discussed in this paper.  It is well known that
this diagonalization requires coordinate rotations, but it also
requires squeeze transformations.  In order to understand fully the
system of two coupled oscillators, we need the group $O(3,3)$ which is
the Lorentz group applicable to a six-dimensional space consisting of
three space-like and three time-like directions~\cite{hkn95jm}.
Indeed, coupled harmonic oscillators and the Lorentz group cannot
be separated from each other.

\begin{appendix}
\section{Contraction of O(3) to E(2)}\label{o3e2}
In this Appendix, we explain what the $E(2)$ group is.  We then
explain how we can obtain this group from the three-dimensional
rotation group by making a flat-surface or cylindrical approximation.
This contraction procedure will give a clue to obtaining the $E(2)$-like
symmetry for massless particles from the $O(3)$-like symmetry for
massive particles by making the infinite-momentum limit.

The $E(2)$ transformations consist of rotation and two translations on
a flat plane.  Let us start with the  rotation matrix applicable to
the column vector $(x, y, 1)$:
\begin{equation}\label{rot}
R(\theta) = \pmatrix{\cos\theta & -\sin\theta & 0 \cr
\sin\theta & \cos\theta & 0 \cr 0 & 0 & 1} .
\end{equation}
Let us then consider the translation matrix:
\begin{equation}
T(a, b) = \pmatrix{1 & 0 & a \cr 0 & 1 & b \cr 0 & 0 & 1} .
\end{equation}
If we take the product $T(a, b) R(\theta)$,
\begin{equation}\label{eucl}
E(a, b, \theta) = T(a, b) R(\theta) =
\pmatrix{\cos\theta & -\sin\theta & a \cr
\sin\theta & \cos\theta & b \cr 0 & 0 & 1} .
\end{equation}
This is the Euclidean transformation matrix applicable to the
two-dimensional $x y$ plane.  The matrices $R(\theta)$ and $T(a,b)$
represent the rotation and translation subgroups respectively.  The
above expression is not a direct product because $R(\theta)$ does not
commute with $T(a,b)$.  The translations constitute an Abelian invariant
subgroup because two different $T$ matrices commute with each other,
and because
\begin{equation}
R(\theta) T(a,b) R^{-1}(\theta) = T(a',b') .
\end{equation}
The rotation subgroup is not invariant because the conjugation
$$T(a,b) R(\theta) T^{-1}(a,b)$$
does not lead to another rotation.

We can write the above transformation matrix in terms of generators.
The rotation is generated by
\begin{equation}
J_{3} = \pmatrix{0 & -i & 0 \cr i & 0 & 0 \cr 0 & 0 & 0} .
\end{equation}
The translations are generated by
\begin{equation}
P_{1} = \pmatrix{0 & 0 & i \cr 0 & 0 & 0 \cr 0 & 0 & 0} , \qquad
P_{2} = \pmatrix{0 & 0 & 0 \cr 0 & 0 & i \cr 0 & 0 & 0} .
\end{equation}
These generators satisfy the commutation relations:
\begin{equation}\label{e2com}
[P_{1}, P_{2}] = 0 , \qquad [J_{3}, P_{1}] = iP_{2} ,
\qquad [J_{3}, P_{2}] = -iP_{1} .
\end{equation}
This $E(2)$ group is not only convenient for illustrating the groups
containing an Abelian invariant subgroup, but also occupies an
important place in constructing representations for the little
group for massless particles, since the little group for massless
particles is locally isomorphic to the above $E(2)$ group.

\begin{figure}[thb]  
\centerline{\psfig{figure=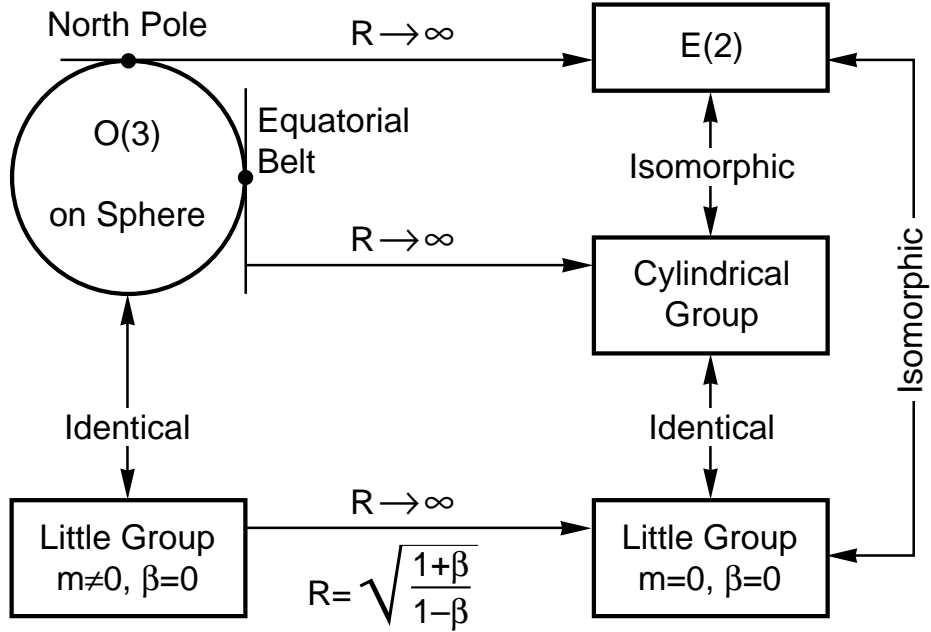,angle=0,height=90mm}}
\vspace{5mm}
\caption{Contraction of O(3) to E(2) and to the cylindrical group,
and contraction of the O(3)-like little group to the E(2)-like
little group.  The correspondence between E(2) and the E(2)-like
little group is isomorphic but not identical.  The cylindrical group
is identical to the E(2)-like little group.  The Lorentz boost of
the O(3)-like little group for a massive particle is the same as
the contraction of O(3) to the cylindrical group.}
\end{figure}

The contraction of $O(3)$ to $E(2)$ is well known and is often called
the Inonu-Wigner contraction~\cite{inonu53}.  The question is whether
the $E(2)$-like little group can be obtained from the $O(3)$-like
little group.  In order to answer this question, let us closely look
at the original form of the Inonu-Wigner contraction.  We start with
the generators of $O(3)$.  The $J_{3}$ matrix is given in Eq.(\ref{j3}),
and
\begin{equation}\label{o3gen}
J_{2} = \pmatrix{0&0&i\cr0&0&0\cr-i&0&0} , \qquad
J_{3} = \pmatrix{0&-i&0\cr i &0&0\cr0&0&0} .
\end{equation}
The Euclidean group $E(2)$ is generated by $J_{3}, P_{1}$ and $P_{2}$,
and their Lie algebra has been discussed in Sec.~\ref{intro}.

Let us transpose the Lie algebra of the $E(2)$ group.  Then $P_{1}$ and
$P_{2}$ become $Q_{1}$ and $Q_{2}$ respectively, where
\begin{equation}
Q_{1} = \pmatrix{0&0&0\cr0&0&0\cr i &0&0} , \qquad
Q_{2} = \pmatrix{0&0&0\cr0&0&0\cr0&i&0} .
\end{equation}
Together with $J_{3}$, these generators satisfy the
same set of commutation relations as that for
$J_{3}, P_{1}$, and $P_{2}$ given in Eq.(\ref{e2com}):
\begin{equation}
[Q_{1}, Q_{2}] = 0 , \qquad [J_{3}, Q_{1}] = iQ_{2} , \qquad
[J_{3}, Q_{2}] = -iQ_{1} .
\end{equation}
These matrices generate transformations of a point on a circular
cylinder.  Rotations around the cylindrical axis are generated by
$J_{3}$.  The matrices $Q_{1}$ and $Q_{2}$ generate translations along
the direction of $z$ axis.  The group generated by these three matrices
is called the {\it cylindrical group}~\cite{kiwi87jm,kiwi90jm}.

We can achieve the contractions to the Euclidean and cylindrical groups
by taking the large-radius limits of
\begin{equation}\label{incont}
P_{1} = {1\over R} B^{-1} J_{2} B ,
\qquad P_{2} = -{1\over R} B^{-1} J_{1} B ,
\end{equation}
and
\begin{equation}\label{cycont}
Q_{1} = -{1\over R}B J_{2}B^{-1} , \qquad
Q_{2} = {1\over R} B J_{1} B^{-1} ,
\end{equation}
where
\begin{equation}\label{bmatrix}
B(R) = \pmatrix{1&0&0\cr0&1&0\cr0&0&R}  .
\end{equation}
The vector spaces to which the above generators are applicable are
$(x, y, z/R)$ and $(x, y, Rz)$ for the Euclidean and cylindrical groups
respectively.  They can be regarded as the north-pole and equatorial-belt
approximations of the spherical surface respectively~\cite{kiwi87jm}.
These two different contraction procedures are illustrated in Fig.~3.

\section{Contraction of O(3)-like Little Group to E(2)-like Little
Group}\label{contrac}

In this appendix, we shall discuss the contraction of $O(3)$-like
little group to the $E(2)$-like little group as the infinite-momentum
or zero-mass limit of the Lorentz-boosted $O(3)$-like little group.
The relation of this procedure to the contraction of $O(3)$ is
illustrated in Fig.~3.

Let us go back to Eq.(\ref{incont}) and Eq.(\ref{cycont}).  Since
$P_{1} (P_{2})$ commutes with $Q_{2} (Q_{1})$, we can consider the
following combination of generators.
\begin{equation}
F_{1} = P_{1} + Q_{1} , \qquad F_{2} = P_{2} + Q_{2} .
\end{equation}
Then these operators also satisfy the commutation relations:
\begin{equation}\label{commuf}
[F_{1}, F_{2}] = 0 , \qquad [J_{3}, F_{1}] = iF_{2} , \qquad
[J_{3}, F_{2}] = -iF_{1} .
\end{equation}
However, we cannot make this addition using the three-by-three matrices
for $P_{i}$ and $Q_{i}$ to construct three-by-three matrices for $F_{1}$
and $F_{2}$, because the vector spaces are different for the $P_{i}$ and
$Q_{i}$ representations.  We can accommodate this difference by creating
two different $z$ coordinates, one with a contracted $z$ and the other
with an expanded $z$, namely $(x, y, Rz, z/R)$.  Then the generators
become
\begin{equation}
P_{1} = \pmatrix{0&0&0&i\cr0&0&0&0\cr0&0&0&0\cr0&0&0&0} , \qquad
P_{2} = \pmatrix{0&0&0&0\cr0&0&0&i\cr0&0&0&0\cr0&0&0&0} .
\end{equation}
\begin{equation}
Q_{1} = \pmatrix{0&0&0&0\cr0&0&0&0\cr i &0&0&0\cr0&0&0&0} , \qquad
Q_{2} = \pmatrix{0&0&0&0\cr0&0&0&0\cr0&i&0&0\cr0&0&0&0} .
\end{equation}
Then $F_{1}$ and $F_{2}$ will take the form
\begin{equation}\label{f1f2}
F_{1} = \pmatrix{0&0&0&i\cr0&0&0&0\cr i &0&0&0\cr0&0&0&0} , \qquad
F_{2} = \pmatrix{0&0&0&0\cr0&0&0&i\cr0&i&0&0\cr0&0&0&0} .
\end{equation}
The rotation generator $J_{3}$ takes the form of Eq.(\ref{j3}).
These four-by-four matrices satisfy the E(2)-like commutation relations
of Eq.(\ref{commuf}).

Now the $B$ matrix of Eq.(\ref{bmatrix}), can be expanded to
\begin{equation}\label{bmatrix2}
B(R) = \pmatrix{1&0&0&0\cr0&1&0&0\cr0&0&R&0\cr0&0&0&1/R} .
\end{equation}
If we make a similarity transformation on the above form using the matrix
\begin{equation}\label{simil}
\pmatrix{1&0&0&0\cr0&1&0&0\cr0&0&1/\sqrt{2} &-1/\sqrt{2}
\cr0&0&1/\sqrt{2}&1/\sqrt{2}} ,
\end{equation}
which performs a 45-degree rotation of the third and fourth coordinates,
then this matrix becomes
\begin{equation}\label{simil2}
\pmatrix{1&0&0&0\cr0&1&0&0\cr0&0 & \cosh\eta & \sinh\eta
\cr0 & 0 & \sinh\eta & \cosh\eta} ,
\end{equation}
with $R = e^\eta$.  This form is the Lorentz boost matrix along the $z$
direction.  If we start with the set of expanded rotation generators
$J_{3}$ of Eq.(\ref{j3}), and
perform the same operation as the original Inonu-Wigner contraction
given in Eq.(\ref{incont}), the result is
\begin{equation}
N_{1} = {1\over R} B^{-1} J_{2} B ,
\qquad N_{2} = -{1\over R} B^{-1} J_{1} B ,
\end{equation}
where $N_{1}$ and $N_{2}$ are given in Eq.(\ref{n1n2}).  The generators
$N_{1}$ and $N_{2}$ are the contracted $J_{2}$ and $J_{1}$ respectively
in the infinite-momentum/zero-mass limit.

This contraction procedure constitutes the second row of Table~I which
contains the expanded content of Einstein's $E = mc^{2}$.  This row
states that the rotational degree of freedom for a massive particle
remains unchanged, and it becomes the helicity degree of freedom
for the massless particle.  The rotational degrees of freedom along
the two transverse direction collapse into one gauge degree of
freedom.

\end{appendix}


\begin{thebibliography}{99}


\bibitem{wig59}
E. P. Wigner, {\it Group Theory and its Application to the Quantum
Mechanics of Atomic Spectra}, translated from the German by J. J.
Griffin, Expanded and Improved Edition (Academic Press, New York,
1959).  For the original version, see E. P. Wigner, {\it Gruppentheorie
und ihre Anwendung auf Quantenmechanik der Atomspektren} (Vieweg,
Braunschweig, 1931).


\bibitem{hofsta55}
R. Hofstadter and R. W. McAllister, Phys. Rev. {\bf 98}, 217 (1955).

\bibitem{gell64}
M. Gell-Mann, Phys. Lett. {\bf 13}, 598 (1964).

\bibitem{owg64}
O. W. Greenberg, Phys. Rev. Lett. {\bf 13}, 598 (1964); O. W. Greenberg
and M. Resnikoff, Phys. Rev. {\bf 163}, 1844 (1967).

\bibitem{fkr71}
R. P. Feynman, M. Kislinger, and F. Ravndal, Phys. Rev. D {\bf 3}, 2706
(1971).

\bibitem{fuji70}
K. Fujimura, T. Kobayashi, and M. Namiki, Prog. Theor. Phys. {\bf 43},
73 (1970).

\bibitem{fey69} 
R. P. Feynman, in {\it High Energy Collisions}, Proceedings of the
Third International Conference, Stony Brook, New York, edited by
C. N. Yang {\it et al.} (Gordon and Breach, New York, 1969);
J. D. Bjorken and E. A. Paschos, Phys. Rev. {\bf 185} 1975 (1969).

\bibitem{wig39}
E. P. Wigner, Ann. Math. {\bf 40}, 149 (1939).

\bibitem{dir45}
P. A. M. Dirac, Proc. Roy. Soc. (London) {\bf A183}, 284 (1945);

\bibitem{yuka53}
H. Yukawa, Phys. Rev. {\bf 91}, 415 (1953).

\bibitem{markov56}
M. Markov, Nuovo Cimento Suppl. {\bf 3} (1956) 760;
V. L. Ginzburgh and V. I. Man'ko, Nucl. Phys. {\bf 74} (1965) 577.

\bibitem{kno79jm}
Y. S. Kim, M. E. Noz, and S. H. Oh, J. Math. Phys. {\bf 20},
1341 (1979).

\bibitem{knp86}
Y. S. Kim and M. E. Noz, {\it Theory and Applications of the Poincar\'e
Group} (Reidel, Dordrecht, 1986).

\bibitem{dir27}
P. A. M. Dirac, Proc. Roy. Soc. (London) {\bf A 114}, 243 and 710
(1927).

\bibitem{fey72}
R. P. Feynman, {\it Statistical Mechanics}
(Benjamin/Cummings, Reading, MA, 1972).

\bibitem{hkn98}
D. Han, Y. S. Kim, and M. E. Noz, {\it Illustrative Example for
Feynman's Rest of the Universe}, Am. J. Phys. (to be published).

\bibitem{inonu53}
E. Inonu and E. P. Wigner, Proc. Natl. Acad. Sci. (U.S.) {\bf 39}, 510
(1953).

\bibitem{marx}
Humboldt University in Berlin has a brilliant history.  In front of
the the main building, there is a statue of Hermann von Helmholtz.
In the lobby of the main hall, there is a marble plate engraved
with a quotation ``Die Philosophen haben dei Weld nur vershieden
interpretiet; es kommt aber darauf an, zie su veraendern.\hspace{1ex}
 -- \hspace{1ex} Karl Marx.''  In English, this quotation could say
``Philosophers interpret this world in various ways. There comes the
question of changing the world.''  It is my recollection that Wigner
never liked Marx, and he did not quote from Marx when he told me about
philosophers.  Yet, what Wigner said coincides with the first half of
what Marx said in the marble plate at Humboldt University.

\bibitem{tani79}
Y. Tanikawa, {\it Hideki Yukawa: Scientific Works} (Iwanami Shoten,
Tokyo, 1979).


\bibitem{apple94}
A. Applebaum, {\it Between East and West, Across the Borderlands of Europe}
(Pantheon Books, New York, 1994)

\bibitem{kim89}
Y. S. Kim, Phys. Rev. Lett. {\bf 63}, 348-351 (1989).

\bibitem{suntzu96}
Sun Tzu and Sun Pin (translated by R. D. Sawyer), {\it The Complete
Art of War} (Westview Press, Boulder, CO, 1996).

\bibitem{wein64}
S. Weinberg, Phys. Rev. {\bf 134}, B882 (1964); {\it ibid.} {\bf 135},
B1049 (1964).

\bibitem{janner71}
A. Janner and T. Janssen,  Physica {\bf 53}, 1 (1971);
{\it ibid.} {\bf 60}, 292 (1972).

\bibitem{kim97poz}
Y. S. Kim, in {\it Symmetry and Structural Properties of Condensed
Matter}, Proceedings 4th International School of Theoretical Physics
(Zajaczkowo, Poland), edited by T. Lulek, W. Florek, and B. Lulek
(World Scientific, 1997).

\bibitem{hks82}
D. Han, Y. S. Kim, and D. Son, Phys. Rev. D {\bf 26}, 3717 (1982).

\bibitem{kim97min}
Y. S. Kim, in {\it Quantum Systems: New Trends and Methods}, Proceedings
of the International Workshop (Minsk, Belarus), edited by Y. S. Kim,
L. M. Tomil'chik, I. D. Feranchuk,  and A. Z. Gazizov (World Scientific,
1997)

\bibitem{ferrara82}
S. Ferrara and C. Savoy, in {\it Supergravity 1981}, S. Ferrara and
J. G. Taylor eds. (Cambridge Univ. Press, Cambridge, 1982), p. 151.
See also P. Kwon and M. Villasante, J. Math. Phys. {\bf 29}, 560 (1988);
{\it ibid.} {\bf 30}, 201 (1989).  For an earlier paper on this subject,
see H. Bacry and N. P. Chang, Ann. Phys. {\bf 47}, 407 (1968).

\bibitem{hks83pl}
D. Han, Y. S. Kim, and D. Son, Phys. Lett. B {\bf 131}, 327 (1983).
See also D. Han, Y. S. Kim, M. E. Noz, and D. Son, Am. J. Phys. {\bf 52},
1037 (1984).

\bibitem{kiwi87jm}
Y. S. Kim and E. P. Wigner, J. Math. Phys. {\bf 28}, 1175 (1987) and
{\bf 32}, 1998 (1991);

\bibitem{kn73}
Y. S. Kim and M. E. Noz, Phys. Rev. D {\bf 8}, 3521 (1973).

\bibitem{knp91}
Y. S. Kim and M. E. Noz, {\it Phase Space Picture of Quantum
Mechanics} (World Scientific, Singapore, 1991).

\bibitem{bj69}
J. D. Bjorken and E. A. Paschos, Phys. Rev. {\bf 185}, 1975 (1969).

\bibitem{hussar81}
P. E. Hussar, Phys. Rev. D {\bf 23}, 2781 (1981).

\bibitem{neu32}
J. von Neumann, {\it Die mathematische Grundlagen der Quanten-mechanik}
(Springer, Berlin, 1932).  See also J. von Neumann, {\it Mathematical
Foundation of Quantum Mechanics} (Princeton University, Princeton, 1955).

\bibitem{wiya63}
E. P. Wigner and M. M. Yanase, {\it Information Contents of Distributions},
Proc. National Academy of Sciences (U.S.A.) {\bf 49}, 910-918 (1963).

\bibitem{fano57}
U. Fano, Rev. Mod. Phys. {\bf 29} 74 (1957).

\bibitem{hkn90pl}
D. Han, Y. S. Kim, and M. E. Noz, Phys. Lett. A {\bf 144} 111 (1990).

\bibitem{ume82}
H. Umezawa, H. Matsumoto, and M. Tachiki, {\it Thermo Field
Dynamics and Condensed States} (North-Holland, Amsterdam, 1982);
A. Mann and M. Revzen, Phys. Lett. {\bf 134A} 273 (1989);
A. K. Ekert and P. L. Knight, Am. J. Phys. {\bf 57} 692 (1989);
A. Mann, M. Revzen, and H. Umezawa, Phys. Lett. {\bf 139A} 197 (1989);
Y. S. Kim and M. Li, Phys. Lett. {\bf 139A} 445 (1989);
A. Mann, M. Revzen, H. Umezawa, Phys. Lett. {\bf 140A} 475 (1989).

\bibitem{hkn97josa}
D. Han, Y. S. Kim, and M. E. Noz, J. Opt. Soc. Am. A {\bf 14}, 2290
(1997); D. Han, Y. S. Kim, and M. E. Noz, Phys. Rev. E {\bf 56}, 6065
(1997).

\bibitem{hkn95jm}
D. Han, Y. S. Kim, and M. E. Noz, J. Math. Phys. {\bf 36}, 3940
(1995).

\bibitem{kiwi90jm}
Y. S. Kim and E. P. Wigner, J. Math. Phys. {\bf 31}, 55 (1990).

\end{thebibliography}
\end{document}